\def\preprint{1}			

\ifdefined\preprint
  \documentclass[preprint,review,12pt]{elsarticle}
\fi
\ifdefined\wordcount
  \documentclass[final,3p,times,twocolumn]{elsarticle}
\fi
\ifdefined\final
  \documentclass[final,3p,times,twocolumn]{elsarticle}
\fi

\usepackage{graphicx,stfloats}
\usepackage{color}

\usepackage{amsmath} 
\usepackage{tikz} 
\usepackage{notoccite} 
\usetikzlibrary{shapes.geometric, arrows}
\usetikzlibrary{datavisualization.formats.functions}
\usetikzlibrary{positioning}

\usepackage{pgfplots}

\usepackage{soul}

\biboptions{sort&compress}

\journal{Proceedings of the Combustion Institute}

\begin{document}

\begin{frontmatter}

\title{The Effect of the Polytropic Index $\gamma$ on the Structure of Gaseous Detonations}

\author[fir]{A. Sow\corref{cor1}}
\ead{asow2@uottawa.ca}
\author[sec]{S. SM. Lau-Chapdelaine}
\author[fir]{M. I. Radulescu}
\ead{matei@uottawa.ca}
\address[fir]{Department of Mechanical Engineering, University of Ottawa, 161 Louis Pasteur, Ottawa, ON, Canada K1N6N5}
\address[sec]{\textcolor{black}{Department of Chemistry and Chemical Engineering, Royal Military College of Canada, 11 Crerar Cres.,Kingston, ON, Canada K7K7B4}}
\cortext[cor1]{Corresponding author:}

\begin{abstract}
The present study aims to clarify the effect of the polytropic index (i.e., the ratio of specific heats in the context of a perfect gas) on the detonation structure. This is addressed by two-dimensional numerical simulations. To ease the clarification of the role of gasdynamics, a simple Arrhenius kinetic law is used  for the chemical model. The activation energy\textcolor{black}{, }normalized by the shock temperature\textcolor{black}{, }  is kept constant to obtain the same reaction rate sensitivity to temperature in all considered mixtures. This procedure  dissociate\textcolor{black}{s} the gasdynamic effects from the chemistry effects. The numerical results reveal that in mixtures with low polytropic ind\textcolor{black}{icies}, the convect\textcolor{black}{ive} mixing is enhanced compared to mixtures with higher polytropic ind\textcolor{black}{icies}. The mixing is evaluated using Lagrangian tracers.   Moreover, mixtures with low polytropic ind\textcolor{black}{icies} are found to have a shorter reaction length than mixtures with high polytropic ind\textcolor{black}{icies}. Also,  \textcolor{black}{for the range of parameters considered in this study} the results indicate that \textcolor{black}{Mach stem} bifurcation in detonations \textcolor{black}{due to jetting} is primarily a gasdynamic driven mechanism.
   
\end{abstract}


\end{frontmatter}

\ifdefined \wordcount
\clearpage
\fi

\section{Introduction}
\label{Introduction}

Self-sustained detonations in gaseous mixtures are almost always unstable with a non-steady three-dimensional cellular structure \cite{lee_detonation_2008,radulescu_failure_2002,austin_role_2003,radulescu_propagation_2003,ng_numerical_2005,radulescu_hydrodynamic_2007}. Generally speaking, detonations can be classified into two categories: \textcolor{black}{detonations with} regular and irregular  structures. The irregularity of a cellular structure is an important indicator \textcolor{black}{of} the propensity of a mixture to detonate. Mixtures with \textcolor{black}{a} regular structure are more difficult to initiate in opposition to mixtures having an irregular structure \cite{radulescu_failure_2002}. 

\begin{figure}[!h]
\centering
\begin{tikzpicture}
%
    \node[above right] (img) at (0,0) {\includegraphics[clip,  scale=0.5,  trim= 0.cm 0.55cm  0cm  0.6cm]{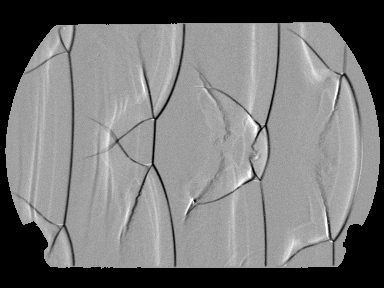} { \includegraphics[clip,  scale=0.51,  trim= 0.cm 0.2cm  0.0cm  0cm]{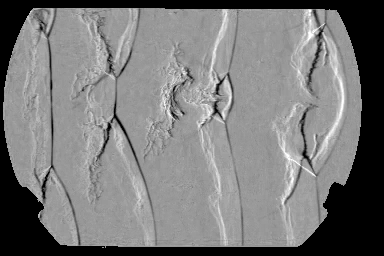}}};
    
  \node at (10pt,120pt) {\textcolor{white}{a}};
      \node at (210pt,120pt) {\textcolor{white}{b}};
%
 \end{tikzpicture}
 \caption{Superimposed schlieren photographs: a): $2\textrm{H}_2 +\textrm{O}_2+7\textrm{Ar}$ at  $4.1$kPa \cite{qxiao_2019}; b): CH$_4$ + $2\textrm{O}_2$ at $3.5$ kPa  \cite{bhattacharjee_experimental_2013}.
}
  \label{fig:qiang}
\end{figure}

For instance, Fig.\  \ref{fig:qiang} illustrates the two-dimensional regular and  irregular cellular structures  \cite{qxiao_2019,bhattacharjee_experimental_2013}. In the regular detonation structure, Fig.\ \ref{fig:qiang}a, as the triple points collide, small forward and backward jets are created and the slip line detaches from the leading front and falls behind it. The detonation appears to be laminar in this configuration with very little small-scale motions  \cite{qxiao_2019}. In more irregular detonations, however, the flow dynamics behind the leading front is much more complex, see Fig.\ \ref{fig:qiang} (b). Pockets of unburned gases are generated after the triple-shocks collision\textcolor{black}{;} the Kelvin-Helmholtz instabilities appear to be more active along the slip line. And, some fine-scale structures with an apparent turbulent motion are observed  \cite{bhattacharjee_experimental_2013}. 

\textcolor{black}{Generation of new cells in detonations has been linked to multiple mechanisms. For weakly unstable detonations like hydrogen-oxygen with argon dilution,  the transverse waves can be reactive. These transverse detonations can host multiple sub-transverse waves that propagate towards the Mach stem. These sub-scale weak pressure waves can reach the main Mach stem and create new entropy waves and triple points, as reported by Asahara {\it{et al}} \cite{asahra2012}. In such configurations, the reactivity of the mixture plays a crucial role on the bifurcation. Moreover, in a double Mach reflection configuration the secondary triple point in the reflected shock can leave traces on the shoot foil and therefore is a potential source of cell multiplication.  The instabilities arising from the coupling taking place between the exothermic heat release and the hydrodynamics have been widely reported as a mechanism of generation of new cells  \cite{manzhalei_fine_1977,gamezo_fine_2000,oran_numerical_1998,jiang_self-organized_2009,austin_reaction_2005,liang_detonation_2007,ng_numerical_2005}. Also, the sensitivity of the reaction rates to temperature perturbations has been pointed out as a mechanism of new cell production \cite{short_cellular_1998}.}

\begin{figure}[!h]
\centering
\begin{tikzpicture}
    \node[above right] (img) at (0,0) {\includegraphics[clip,  scale=0.6,  trim= 0.4cm 12.8cm  0.4cm  7.0cm]{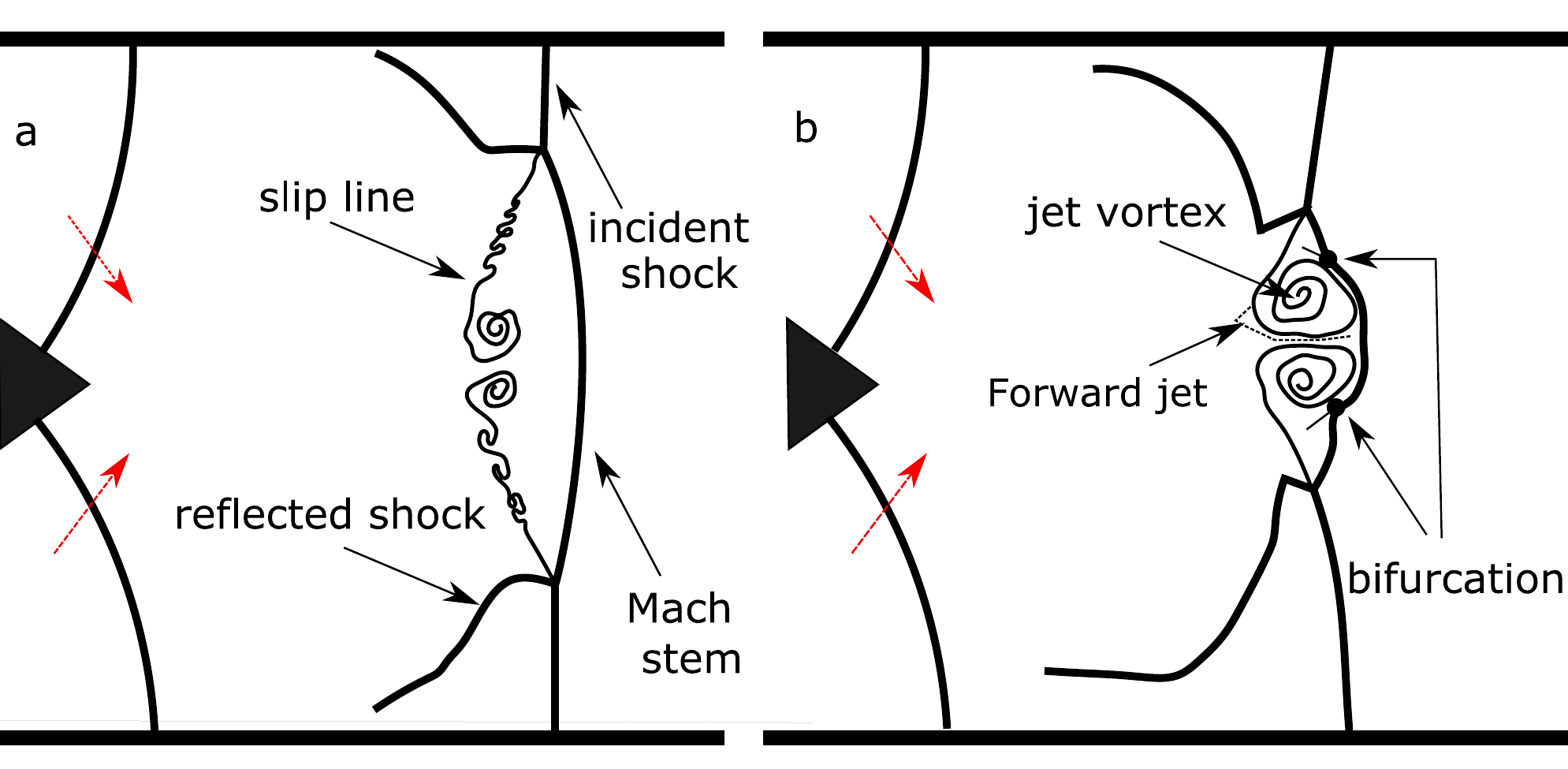}};        
 \end{tikzpicture}
 \caption{\textcolor{black}{Sketch illustrating Mach stem bifurcation from jetting. a) High gamma case ; b) Low gamma case. 
 The dotted arrows indicate the propagation direction of the shock.}
}
  \label{fig:jetting}
\end{figure}

\textcolor{black}{In more unstable detonations, the transverse waves are generally non-reactive and pockets of unburnt gas accumulate behind the front \cite {subbotin1975two}.  These pockets burn by turbulent mixing \cite{radulescu_hydrodynamic_2007,radulescu_detonation_2018}.  It has been recently proposed that triple-shock collisions in these unstable mixtures generate both rear-facing and front-facing jets, which are responsible for promoting turbulent mixing in the reaction zone \cite{lau-chapdelaine_viscous_2019}.  In the present study, we wish to evaluate the influence of the specific heat ratio on these jetting structures shown qualitatively in Fig.\ \ref{fig:jetting}.}

\textcolor{black}{The importance of the specific heat ratio is borne out from the calculations of Mach \cite{philipmachthesis2011} and Lau-Chapdelaine \cite{lau-chapdelaine_viscous_2019}.  The latter authors investigated inert shock reflections. They showed that for high ratios of specific heats and low incident Mach numbers   the forward jets did not bifurcate the Mach stem,  Fig.\ \ref{fig:jetting}a. For low ratios of specific heats and high incident Mach numbers, however, the forward jets generate new triple points on the Mach stem, Fig.\ \ref{fig:jetting}b.
\\
Mach and Radulescu \cite{mach_mach_2011}  have also proposed that the propensity for forward jets may control the birth of new cells.  They found out that conditions in which forward jets were prevalent in inert reflections correlated very well with the experimentally observed irregularity of cellular detonations. }

\textcolor{black}{In the present paper, we wish to investigate the effect of the ratio of specific heats in detonation wave structure.  We wish to determine how  the forward (and rear) jetting observed in inert Mach reflections control the structure of reactive gaseous detonations.}

From an experimental view point, it is extremely difficult to accurately probe the flowfield due to not only the high spatial and temporal resolution requirements but also to the difficulty to isolate the gasdynamics contribution from the chemical contribution. The strategy adopted in this study is to conduct a series of highly resolved numerical simulations of the detonation cellular structure utilizing mixtures with different polytropic ind\textcolor{black}{icies} but sharing the same activation energy, with respect to the post-shock state. Therefore, all mixtures considered in this work have the same reaction rate sensitivity to temperature allowing to differentiate between the gasdynamics contribution and the chemical contribution. 
       
\section{Mathematical and Numerical Details}
\label{mathandnum}
 In this study, we focus mainly on the meso-scale convective effects. Thus, the non-dimensional reactive Euler equations for a calorically perfect gas described below are adopted:
\begin{align} 
\nonumber
\displaystyle \frac{D\rho}{D  t}+ \rho \nabla. U   = 0 \;\;;\;\; 
\displaystyle \rho\frac{D \ U}{D t}+  \nabla p    = 0\;; \\
\displaystyle \rho\frac{D e}{D  t} = \frac{p}{ \rho} \frac{ D \rho}{D t}  +  \rho  Q \frac{D \lambda}{D t} \;\;; \;\;
\displaystyle  \frac{D \lambda}{D  t} = - k  \lambda exp({- E_a/R T}) 
\end{align}

\noindent where $D/D  t$ is the material derivative, $ p$  the pressure, $\rho$  the density, $\lambda$ the reaction progress variable (0 in the products and 1 in the reactants) and $Q$ the heat release. $E_a$ is the activation energy, $\displaystyle  e=  R T/( \gamma -1)$ is the internal energy per unit mass while $\displaystyle \gamma$ is the polytropic index and  $ R$ is the specific gas constant.  The ideal gas law is used to obtain the temperature $\displaystyle T= p/ \rho   R$. 
Nondimensionalization uses $\tilde \rho_0, \; \tilde p_0$,  and $ \tilde \Delta_{1/2}$ as characteristic scales. The parameters $Q$ and $E_a$ are defined here as $Q=\tilde Q/R \tilde T_0$ and  $E_a=\tilde E_a/R \tilde T_0$, respectively.  The tilde indicates a dimensional variable and the subscript $0$ refers to the initial state.  The pre-exponential factor $k$ is chosen so that the  reference length scale $\Delta_{1/2}$,  the distance at which half of the reactant is consumed in the steady Zel'dovich-von Neumann-Döring (ZND) wave, is scaled to unity length. The nondimensional equations are solved using mg, a second-order Gudonov solver with adaptive mesh refinement capabilities (AMR)   \cite{falle_self-similar_1991,falle_upwind_1996}.   A two-dimensional channel, $600 \Delta_{1/2}\times 5 \Delta_{1/2}$,  is considered here. The  detonation propagates from the left to the right in the positive $x$-direction. Reflective boundary conditions are imposed to the top and bottom sides and free boundary conditions are imposed to the left and right sides of the channel, although these do not affect the results presented as they are not in the domain of dependence of the reaction zone structure. The computations are initialized using a ZND solution placed $100  \Delta_{1/2}$ away from the left boundary.  An initial disturbance is added in the density of the fresh mixture to induce appearance of a single transverse wave and the formation of half a cell \cite{sharpe_transverse_2001,mahmoudi_triple_2012,mazaheri_diffusion_2012}  as follows:
 \begin{equation} 
\rho ' =  \left \{
\begin{array}{lrr} 
\nonumber
 0  &,   x < 0   \\
  0.25 [1+cos(\pi y /L)]sin(\pi(1-x)) &,  0\leq x \leq 1   \\
 0   &,   x > 1,  
\end{array}
 \right. 
\end{equation}
          
\noindent where $L$ is the domain width. To guarantee that the final solution is free from the initial perturbation and a stationary solution is established, the detonation is allowed to propagate more than $10$ cell length\textcolor{black}{s}, $200 \times \Delta_{1/2}$.
Unless otherwise stated a base mesh of 1 point per half-reaction zone and 8 \textcolor{black}{refinement levels} are used. \textcolor{black}{ The AMR uses hierarchical series of rectangular cartesian grids. Level 0 and 1 cover all the domain initially. If the local changes of density between existing grid levels is higher than $1\%$ or the local changes in $\rho \lambda$ is higher $0.1\%$ then the cell needs to be refined. Specifying a based grid of 1 point per half-ration zone will give an effective base grid of 2 points per half-reaction length at level 1. The resolution is doubled between two consecutive levels (from level 1 to level 2, the resolution increases from 2 to 4 points per half-reaction length). The maximum resolution is given by $2^{\textrm{maximum  level -1}} $. Readers can find more details on the AMR in \cite{falle_upwind_1996,maxwell2017influence}}.  Thus, regions of interest (shocks, reaction zone) are covered with a fine mesh resolution of $128$ points per half-reaction length.

\section {Detonation structure}
\label{results}

 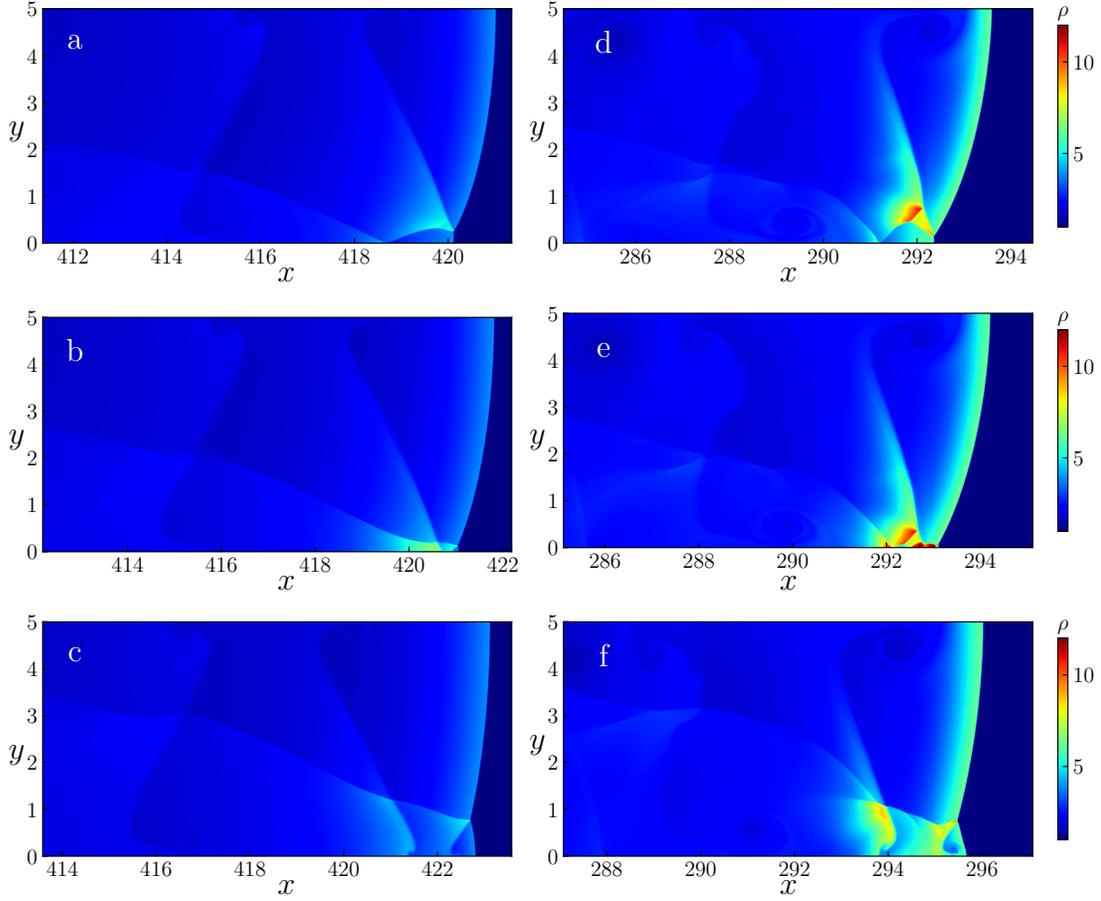
\begin{figure*}[!h]
\centering
\begin{tikzpicture}
  \node[above right] (img) at (0,0) {\includegraphics[clip,  scale=0.55,  trim= 0.65cm 2.9cm  1.8cm  3.3cm]{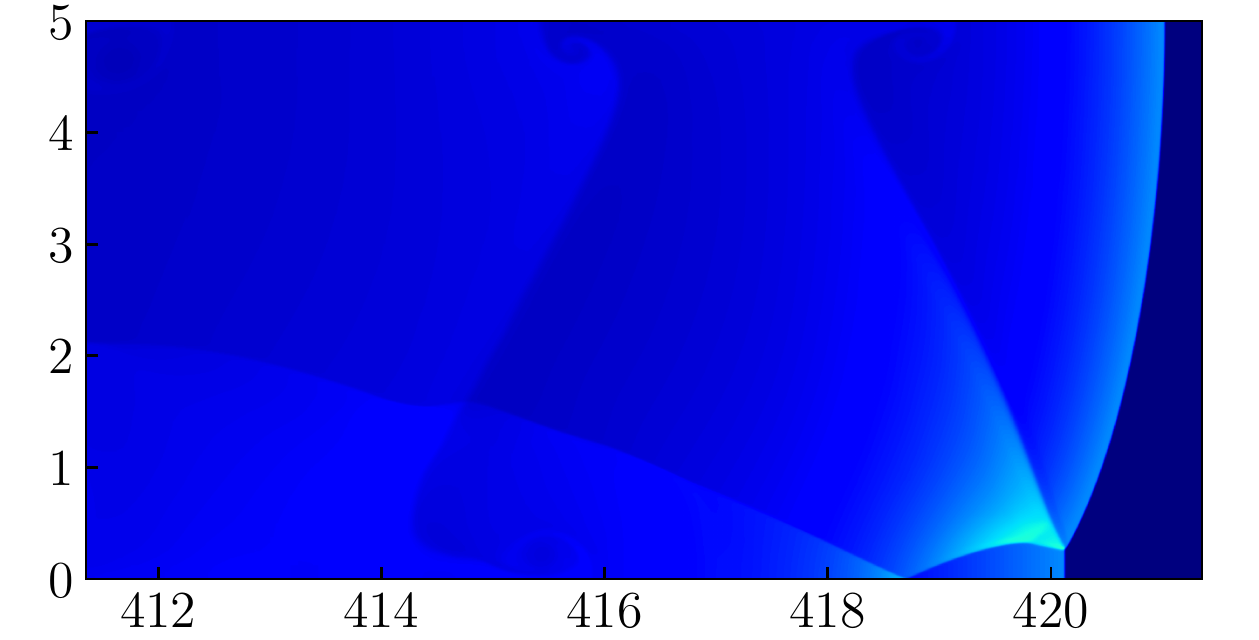} \includegraphics[clip,  scale=0.55,  trim= 1.1cm 2.9cm  0.cm  3.3cm]{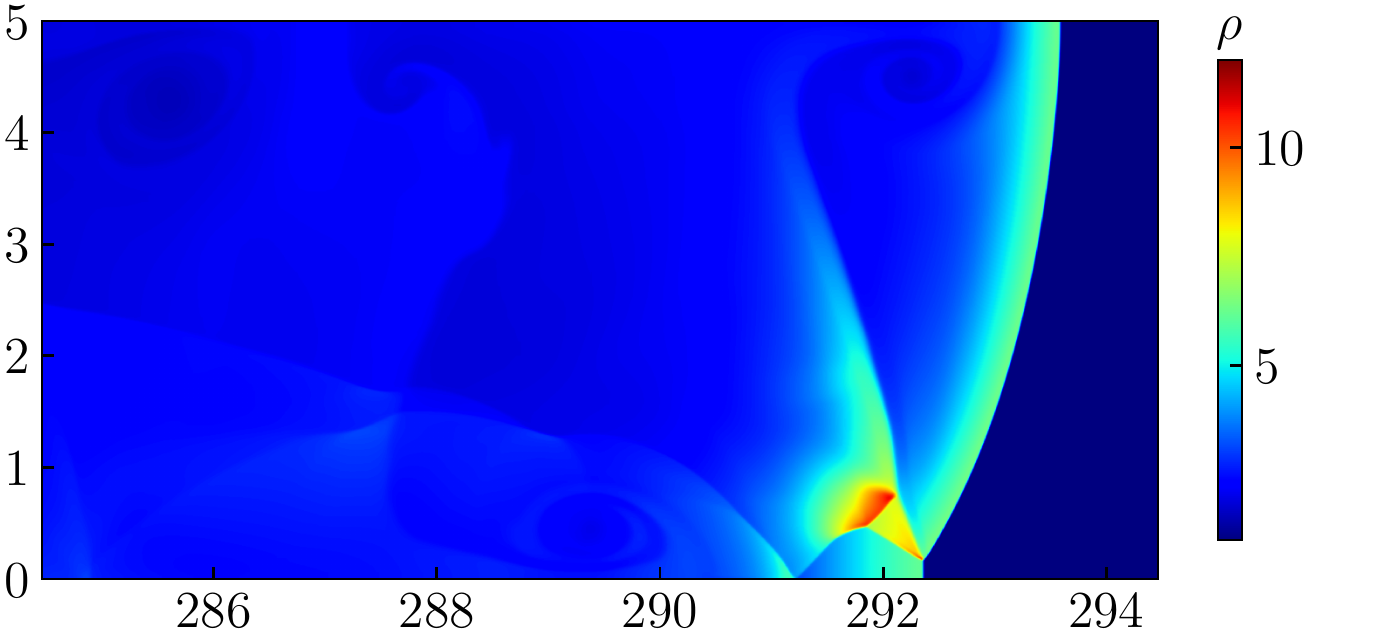}};
  \node at (30pt,90pt) {\textcolor{white}{a}};
        \node at (230pt,90pt) {\textcolor{white}{d}}; 
          \node at (8pt,56pt) {\textcolor{black}{$y$}};
       \node at (110pt,2pt) {\textcolor{black}{$x$}};
                        \node at (205pt,56pt) {\textcolor{black}{$y$}};
       \node at (300pt,2pt) {\textcolor{black}{$x$}};
 \end{tikzpicture}
  \begin{tikzpicture}
  \node[above right ] (img) at (0,0) {\includegraphics[clip,  scale=0.55,  trim= 0.65cm 2.9cm  1.8cm  3.3cm]{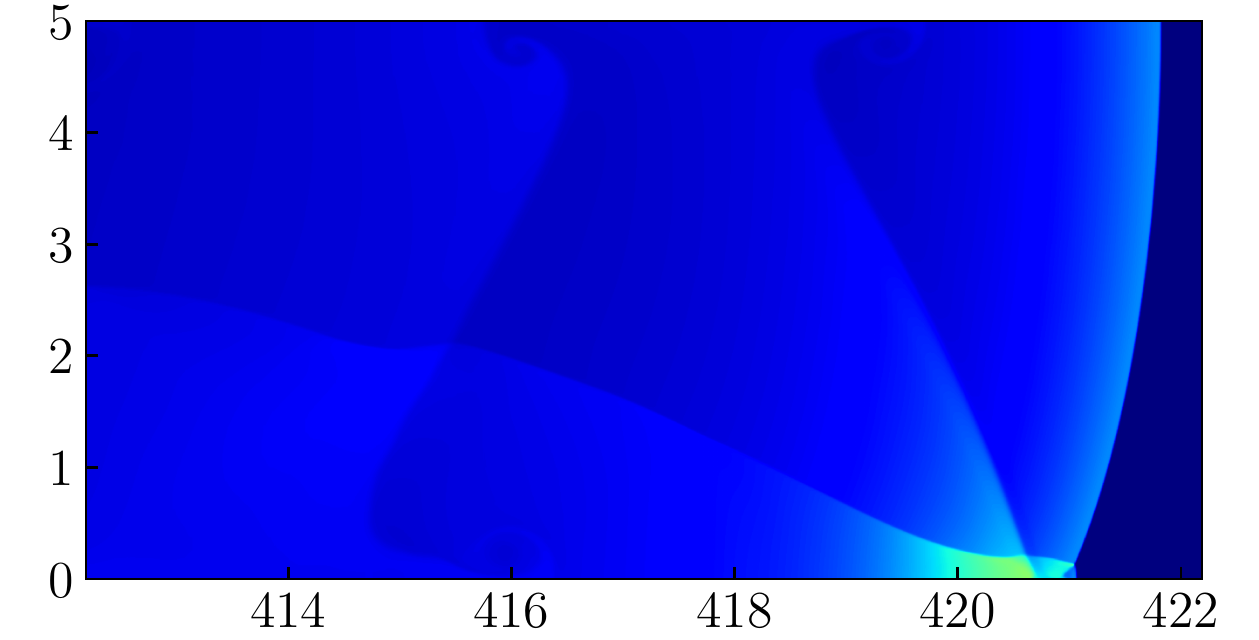}  \includegraphics[clip,  scale=0.55,  trim= 1.1cm 2.8cm  0cm  3.3cm]{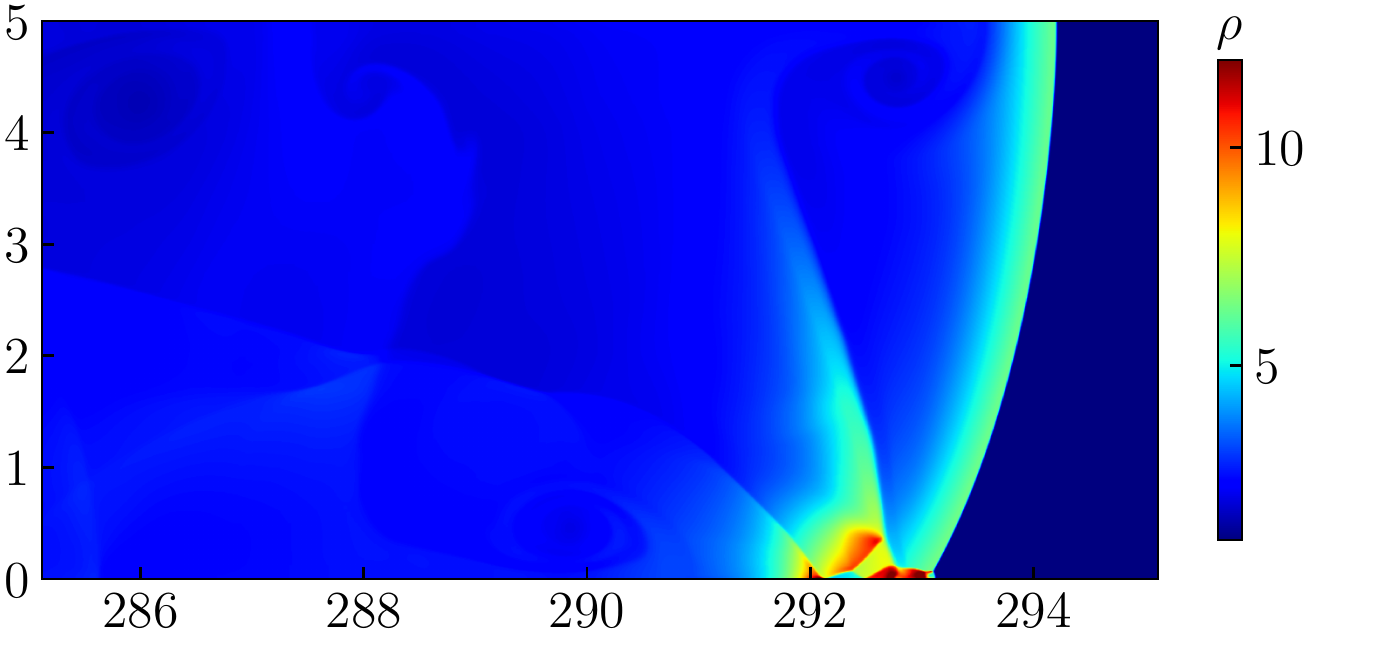} };
  \node at (30pt,90pt) {\textcolor{white}{b}};
        \node at (230pt,90pt) {\textcolor{white}{e}};
          \node at (8pt,56pt) {\textcolor{black}{$y$}};
       \node at (110pt,2pt) {\textcolor{black}{$x$}};
                 \node at (205pt,56pt) {\textcolor{black}{$y$}};
       \node at (300pt,2pt) {\textcolor{black}{$x$}};
       
    \end{tikzpicture}
    
    \begin{tikzpicture}
  \node[above right] (img) at (0,0) {\includegraphics[clip,  scale=0.55,  trim= 0.65cm 2.9cm  1.8cm  3.3cm]{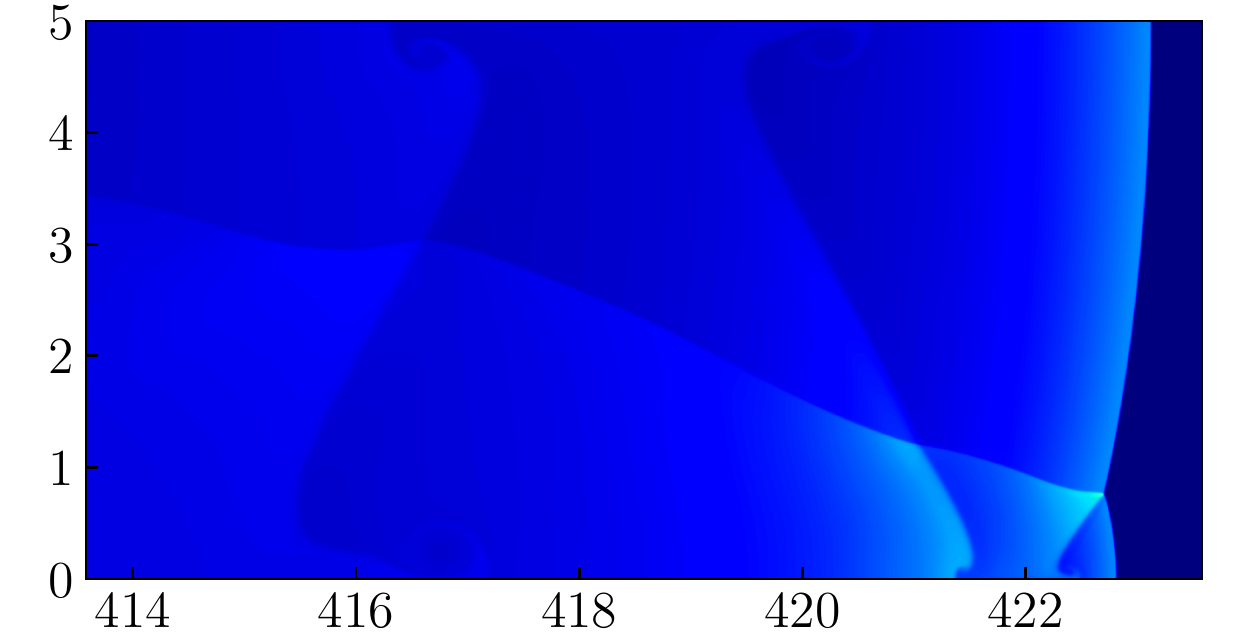} \includegraphics[clip,  scale=0.55,  trim= 1.1cm 2.9cm  0.cm  3.3cm]{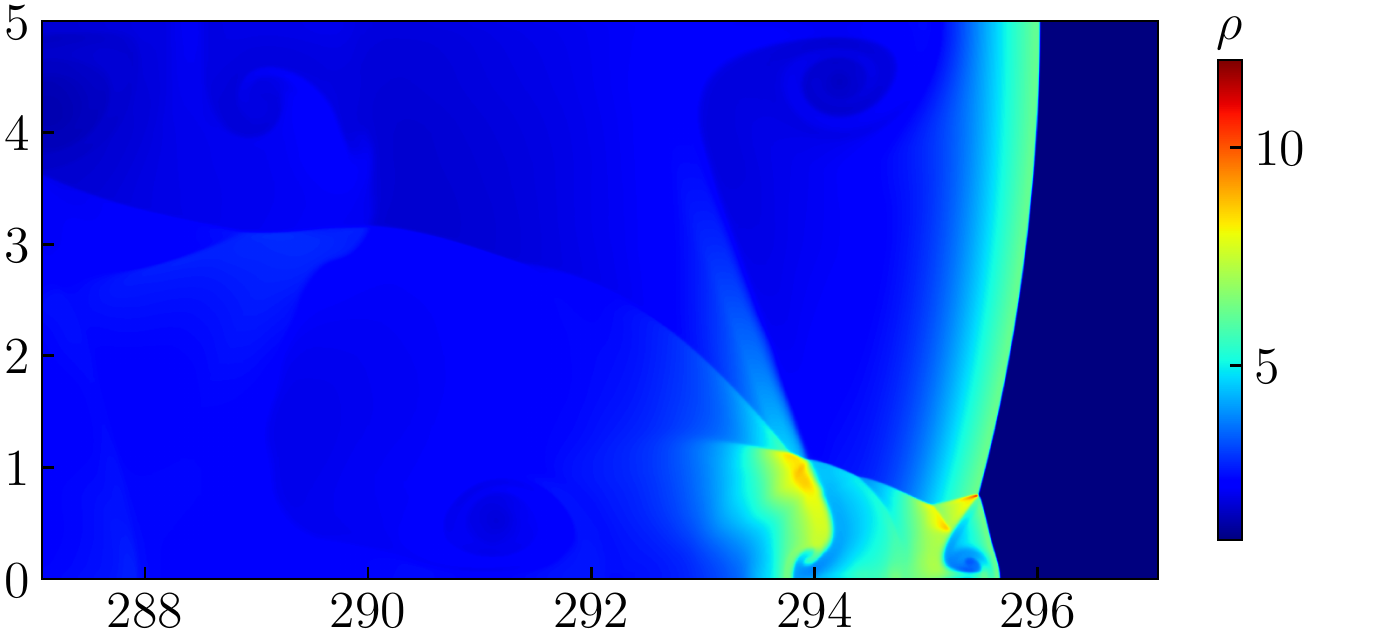}};
  \node at (30pt,90pt) {\textcolor{white}{c}};
        \node at (230pt,90pt) {\textcolor{white}{f}};
                  \node at (8pt,52pt) {\textcolor{black}{$y$}};
       \node at (110pt,2pt) {\textcolor{black}{$x$}};
                        \node at (205pt,56pt) {\textcolor{black}{$y$}};
       \node at (300pt,2pt) {\textcolor{black}{$x$}};
    \end{tikzpicture}  
    
\caption{Contour of density for: $\gamma=1.66$a-c  and  $\gamma=1.33$d-f. Time sequences from  \textcolor{black}{a-c} are:  $t=31.42$, $t=31.49$ , and $t=31.7$.  Time sequences from  \textcolor{black}{d-f} : $t=30.2$, $t=30.26$ , and $t=30.36$.   }
  \label{fig:q50-g-1.66-vs-1.33}
\end{figure*}

The parameters of the numerical experiments conducted in this section are  summarized in table \ref{table:1}.
Figure \ref{fig:q50-g-1.66-vs-1.33} shows the evolution of the obtained flowfield at different times for $\gamma=1.66$ and $\gamma=1.33$. The time sequence starts when the triple-shock is about to reflect from the bottom symmetrical wall, see Fig.\ \ref{fig:q50-g-1.66-vs-1.33}a and Fig.\ \ref{fig:q50-g-1.66-vs-1.33}d for $\gamma=1.66$ and $\gamma=1.33$,  respectively.  For the highest polytropic index considered, $\gamma=1.66$, upon collision the slip line detaches from the leading front, see Fig.\ \ref{fig:q50-g-1.66-vs-1.33}b. A very weak vortex\textcolor{black}{-}like structure is formed at the edge of the slip line attached to the shock. The vortex\textcolor{black}{-}like structures associated \textcolor{black}{with} the detached slip line grow slowly as time evolves.  There is no noticeable forward jetting observed as shown in Fig.\ \ref{fig:q50-g-1.66-vs-1.33}a-c.  

For the intermediate polytropic index  $\gamma=1.33$, as the triple-shock reflects from the bottom symmetrical wall, a higher pressure region (compared to the case with \mbox{$\gamma=1.66$}) is generated behind the main front, see Fig.\ \ref{fig:q50-g-1.66-vs-1.33}e. Figure \ref{fig:q50-g-1.66-vs-1.33}\textcolor{black}{f}  clearly shows the forward jet that bulges the Mach stem. Also, a noticeable difference between $\gamma=1.66$ and $\gamma=1.33$ is the switch \textcolor{black}{from} a transitional Mach reflection (Fig.\ \ref{fig:q50-g-1.66-vs-1.33}c)  \textcolor{black}{to} a double Mach reflection (Fig.\ \ref{fig:q50-g-1.66-vs-1.33}f).   
Moreover, the vortex-like structures noticed for $\gamma=1.66$ are now more active.  Note that the bulge in Fig.\  \ref{fig:q50-g-1.66-vs-1.33}\textcolor{black}{f} disappears later on very quickly as the reflected triple-shock moves toward the upper wall.   

\begin{table}
\centering 
\begin{tabular}{l*{6}{c}r}
\hline
$\gamma $  &  1.1  & 1.2 &1.33 & 1.66  \\
\hline
$E_a/RT_s$   & 4.155      & 4.155 & 4.155 & 4.155   \\

$Q/RT_0$  & 50      & 50 & 50 & 50   \\


\hline
\end{tabular}
\caption{Simulations parameters for different mixtures. }
\label{table:1}
\end{table}

%
%
%

\begin{figure*}[!h]
\centering
\begin{tikzpicture}
  \node[above right] (img) at (0,0) {\includegraphics[clip,  scale=0.55,  trim= 0.65cm 2.8cm  0.cm  3.3cm]{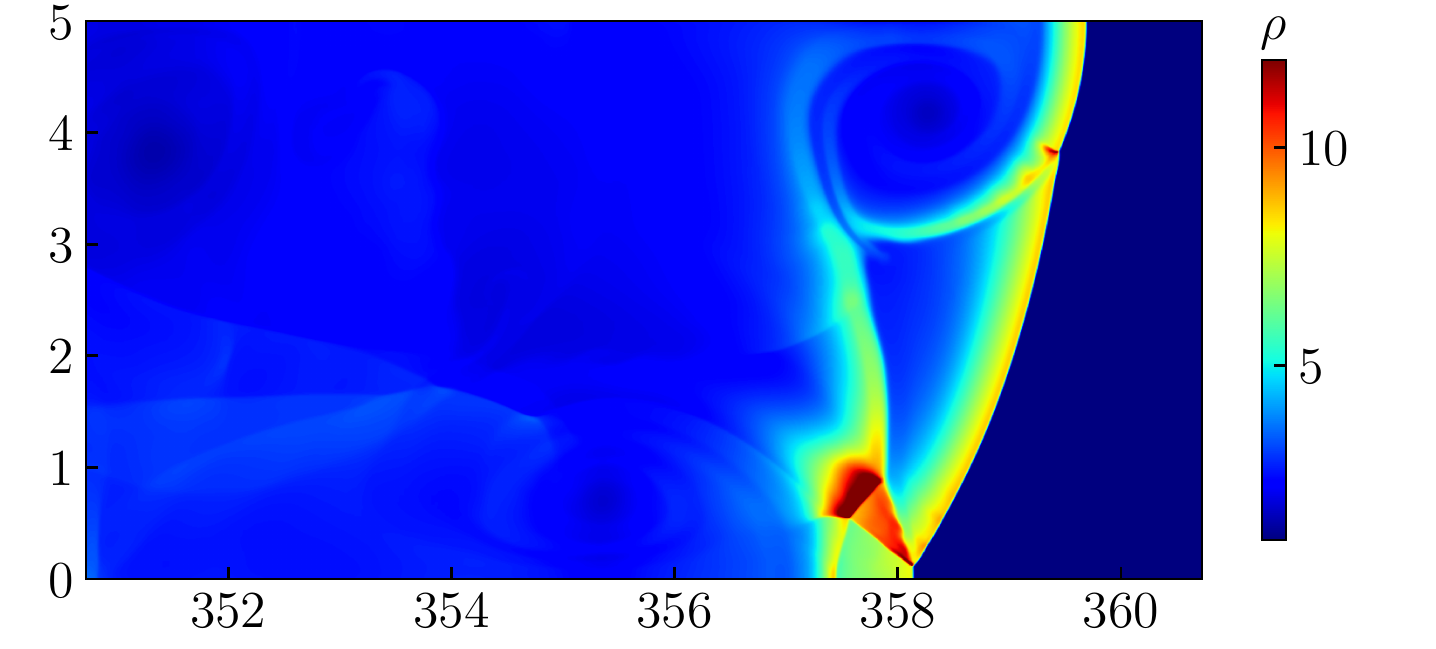} \includegraphics[clip,  scale=0.55,  trim= 0.65cm 2.8cm  0.cm  3.3cm]{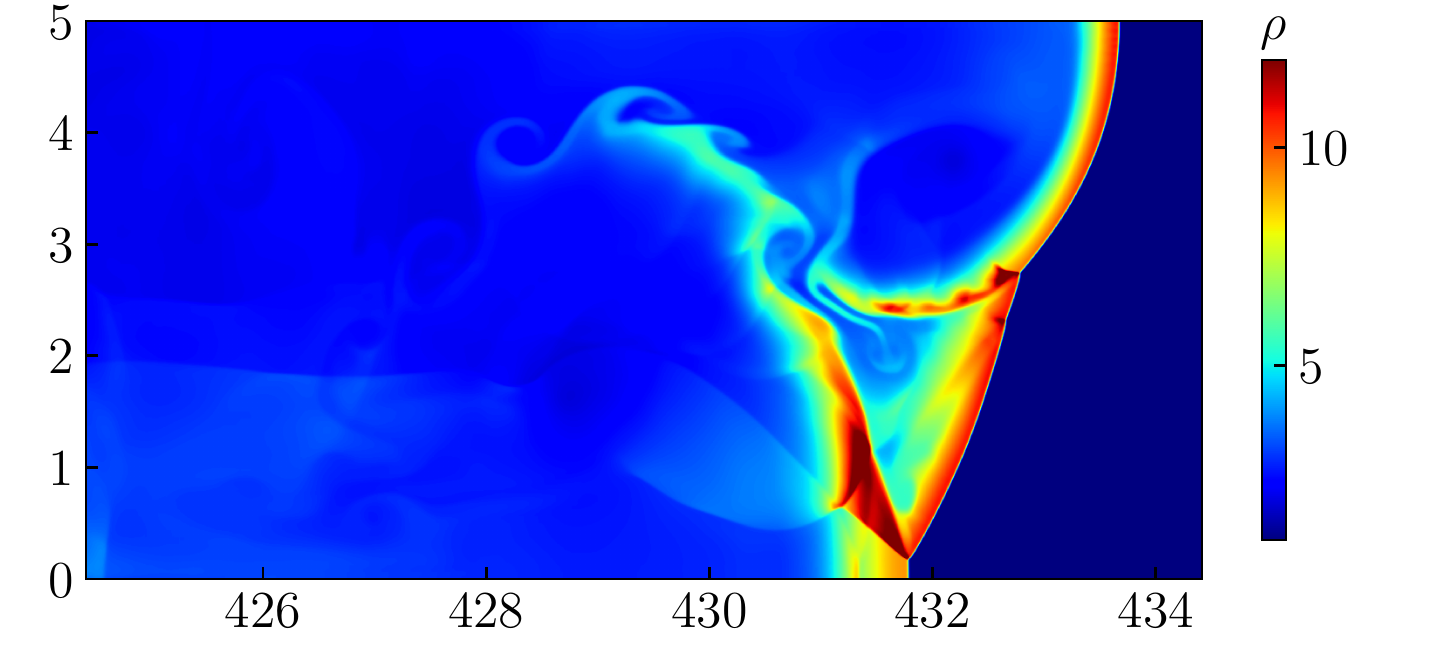}};
  \node at (30pt,90pt) {\textcolor{white}{a}};
        \node at (260pt,90pt) {\textcolor{white}{e}};
                  \node at (8pt,56pt) {\textcolor{black}{$y$}};
       \node at (110pt,2pt) {\textcolor{black}{$x$}};
                         \node at (238pt,56pt) {\textcolor{black}{$y$}};
       \node at (340pt,2pt) {\textcolor{black}{$x$}};

   \draw[white, -latex,thick ](3.8,3.) -- (4.9 ,3.2);
   \node at (3.2,3.) {\textcolor{white}{vortex}};  
   
      \draw[white, -latex,thick ](6.5,2.) -- (6.05 ,2.9);
   \node at (6.5,1.8) {\textcolor{white}{k$1$}};
      \draw[white, -latex,thick ](6.,1.2) -- (5.25 ,0.6);
   \node at (6.25,1.3) {\textcolor{white}{tp}};

 \end{tikzpicture}
 
\begin{tikzpicture}
  \node[above right ] (img) at (0,0) {\includegraphics[clip,  scale=0.55,  trim= 0.65cm 2.8cm  0cm  3.3cm]{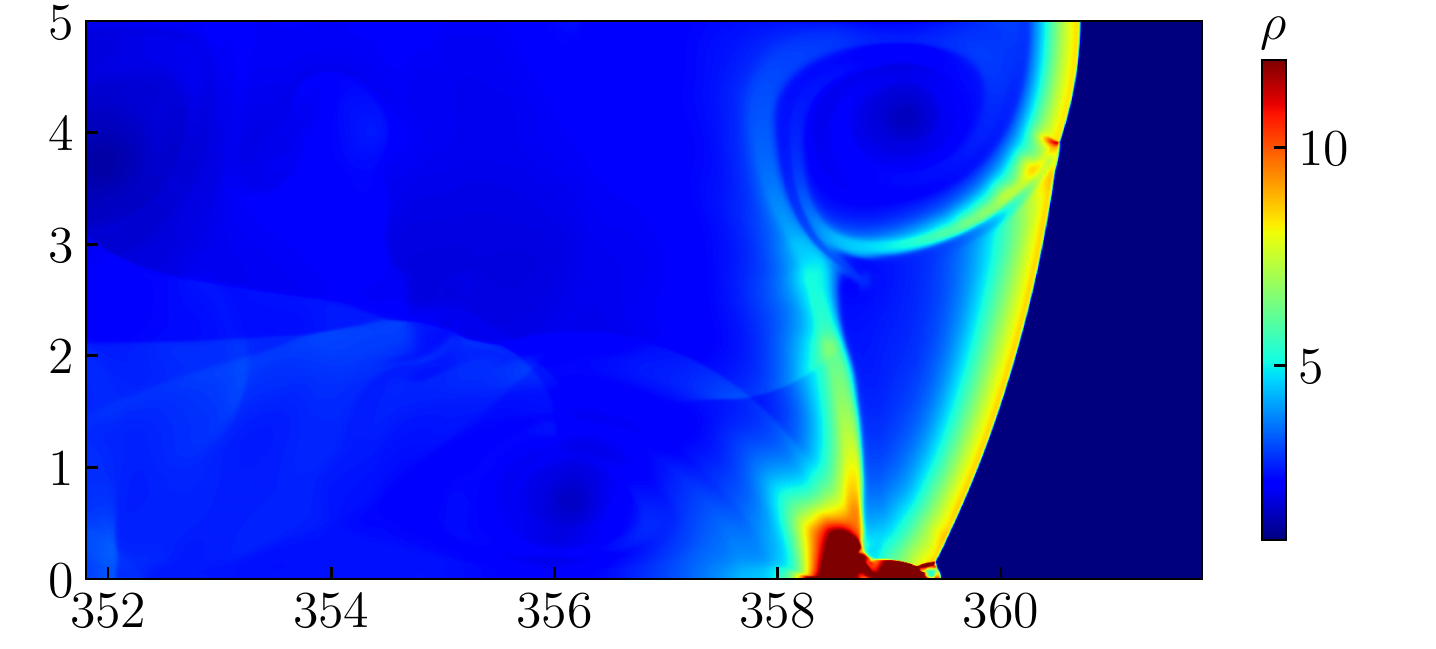}  \includegraphics[clip,  scale=0.55,  trim= 0.65cm 2.8cm  0cm  3.3cm]{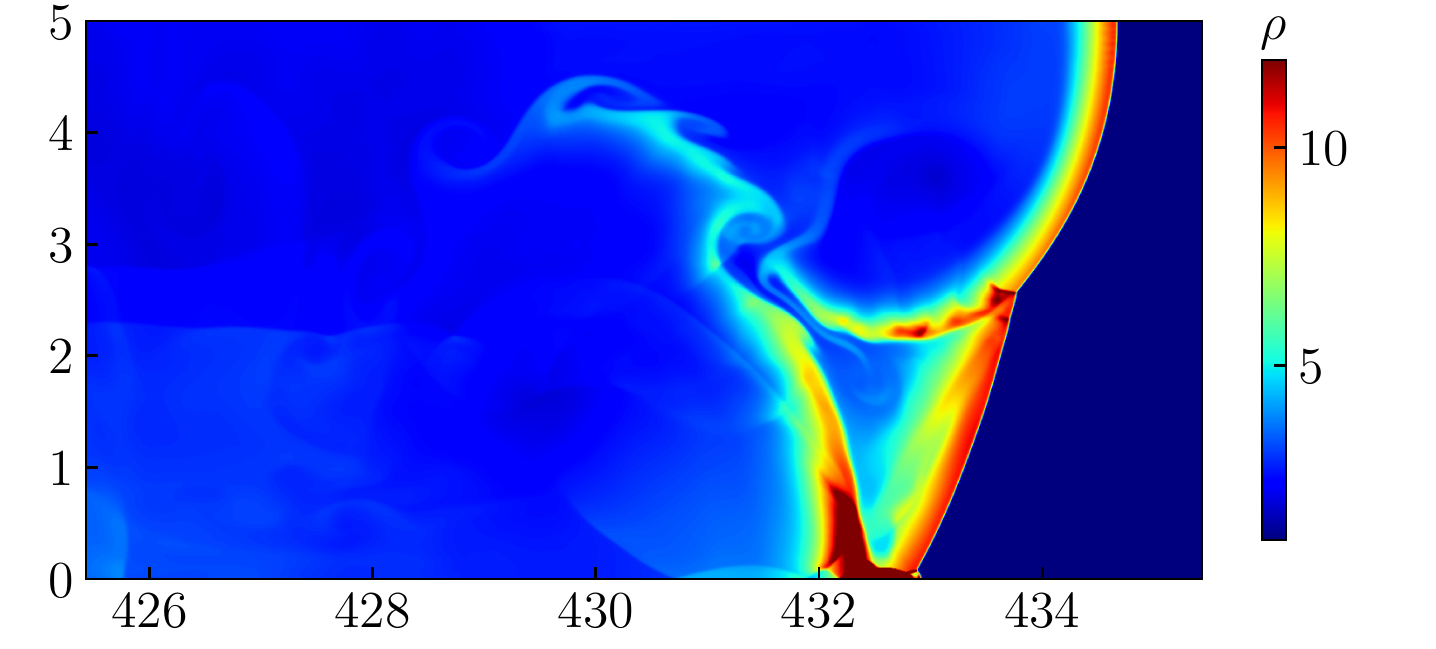} };
    \node at (30pt,90pt) {\textcolor{white}{b}};
        \node at (260pt,90pt) {\textcolor{white}{f}};
           \node at (2.6,1.1) {\textcolor{white}{compression}};
              \node at (2.6,0.8) {\textcolor{white}{wave}};
              \draw[white, -latex,thick ](2.7,1.3) -- (3.5,1.9);
                        \node at (8pt,56pt) {\textcolor{black}{$y$}};
       \node at (110pt,2pt) {\textcolor{black}{$x$}};
                               \node at (238pt,56pt) {\textcolor{black}{$y$}};
       \node at (340pt,2pt) {\textcolor{black}{$x$}};
       
\end{tikzpicture}
    
\begin{tikzpicture}
  \node[above right] (img) at (0,0) {\includegraphics[clip,  scale=0.55,  trim= 0.65cm 2.9cm  0.cm  3.3cm]{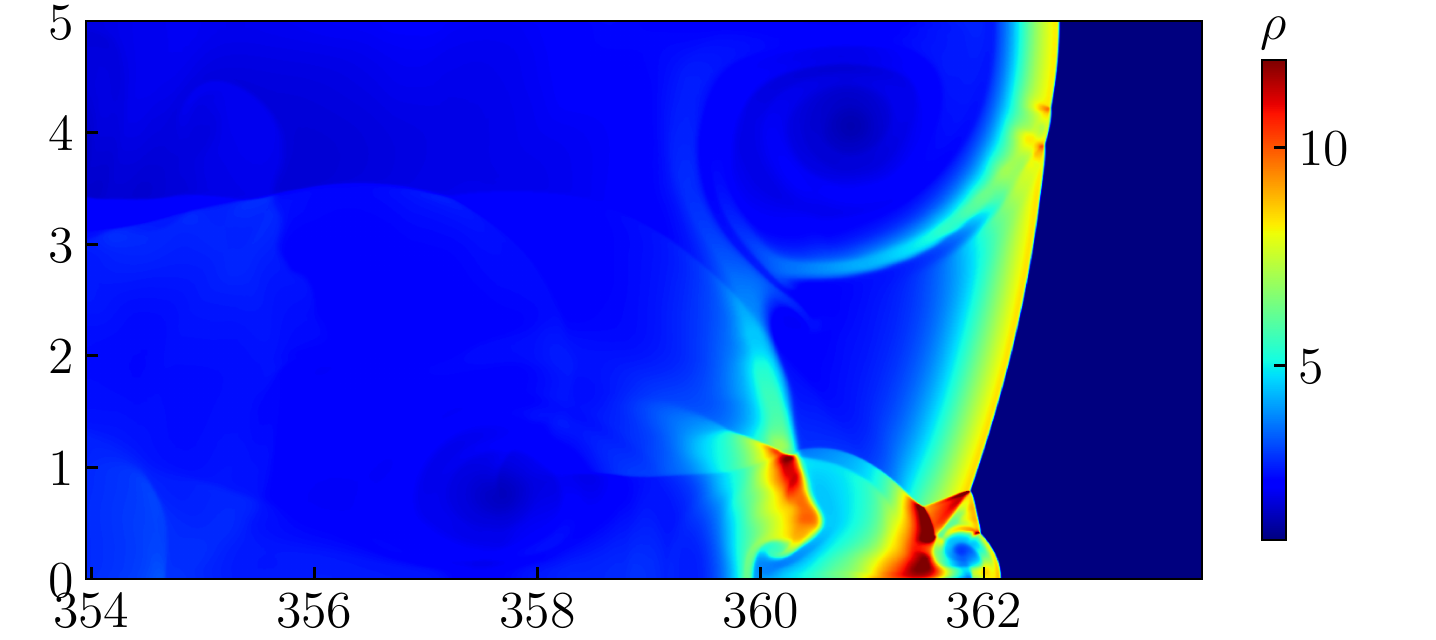} \includegraphics[clip,  scale=0.55,  trim= 0.65cm 2.9cm  0.cm  3.3cm]{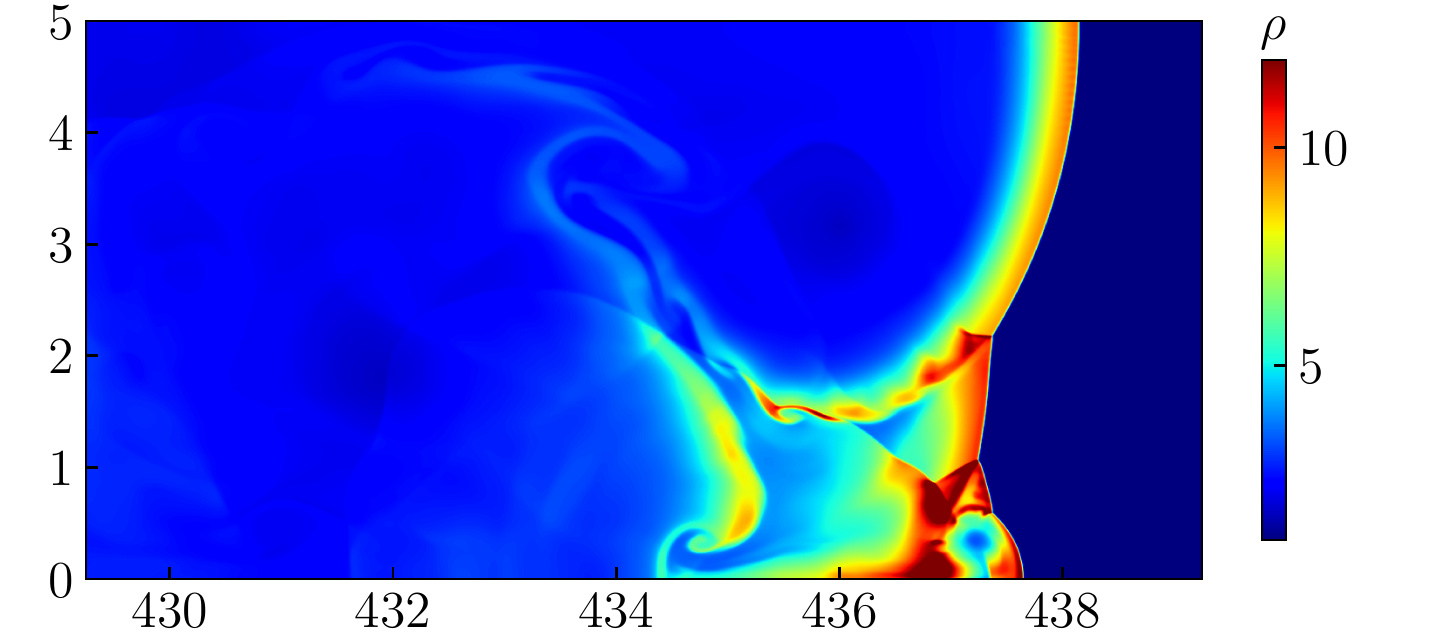}};
    \node at (30pt,90pt) {\textcolor{white}{c}};
        \node at (260pt,90pt) {\textcolor{white}{g}};
           \node at (6.5,1.85)  {\textcolor{white}{rtp}};
             \draw[white, -latex,thick ](6.4,1.7) -- (5.6 ,1.);

            \node at (6.5,1.23)  {\textcolor{white}{k2}};
             \draw[white, -latex,thick ](6.3,1.2) -- (5.6 ,0.71);

              \node at (14.6,2.9)  {\textcolor{white}{k1}};
                \draw[white, -latex,thick ](14.6,2.7) -- (13.8 ,1.8);

                \node at (14.7,1.22)  {\textcolor{white}{rtp}};
                \draw[white, -latex,thick ](14.4,1.2) -- (13.7 ,1.15);

                 \node at (14.7,0.7)  {\textcolor{white}{k2}};
             \draw[white, -latex,thick ](14.4,0.6) -- (13.8 ,0.82);

                           \node at (8pt,56pt) {\textcolor{black}{$y$}};
       \node at (110pt,2pt) {\textcolor{black}{$x$}};
                                  \node at (238pt,56pt) {\textcolor{black}{$y$}};
       \node at (340pt,2pt) {\textcolor{black}{$x$}};
       
\end{tikzpicture}
    
 \begin{tikzpicture}
  \node[above right ] (img) at (0,0) {\includegraphics[clip,  scale=0.55,  trim= 0.65cm 2.8cm  0cm  3.3cm]{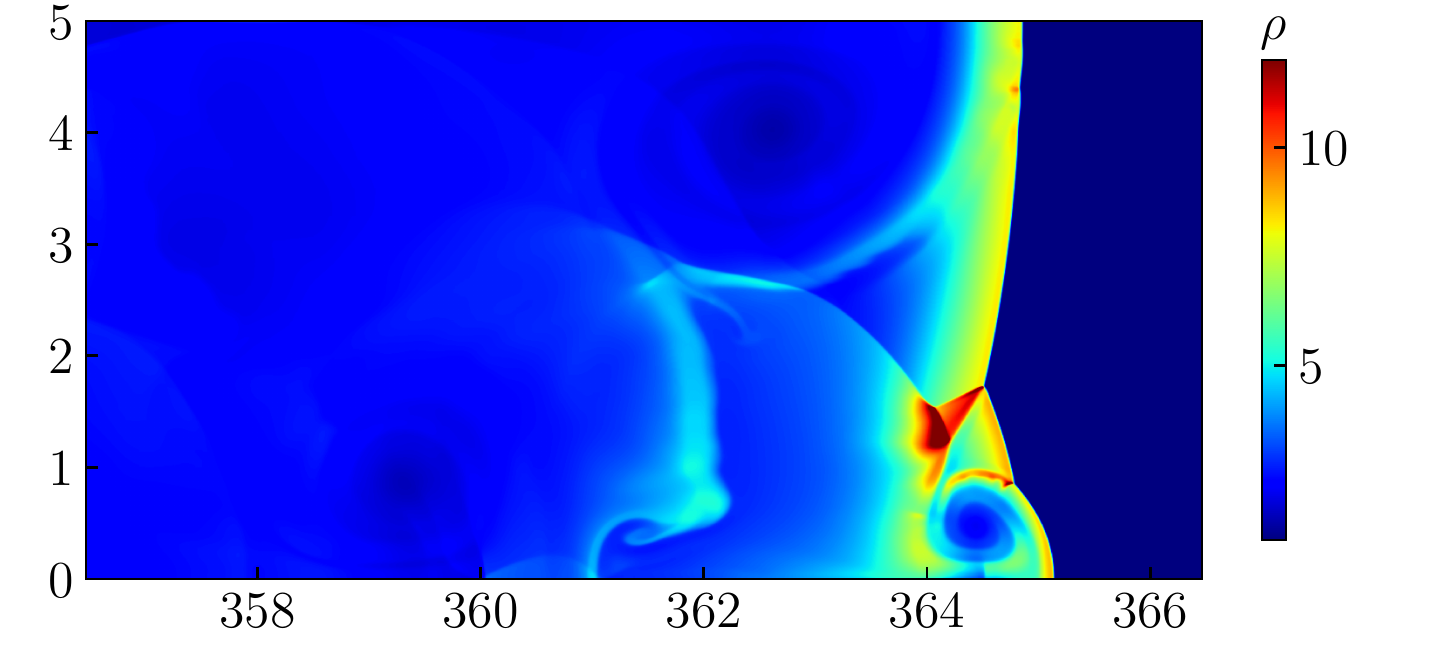}  \includegraphics[clip,  scale=0.55,  trim= 0.65cm 2.8cm  0cm  3.3cm]{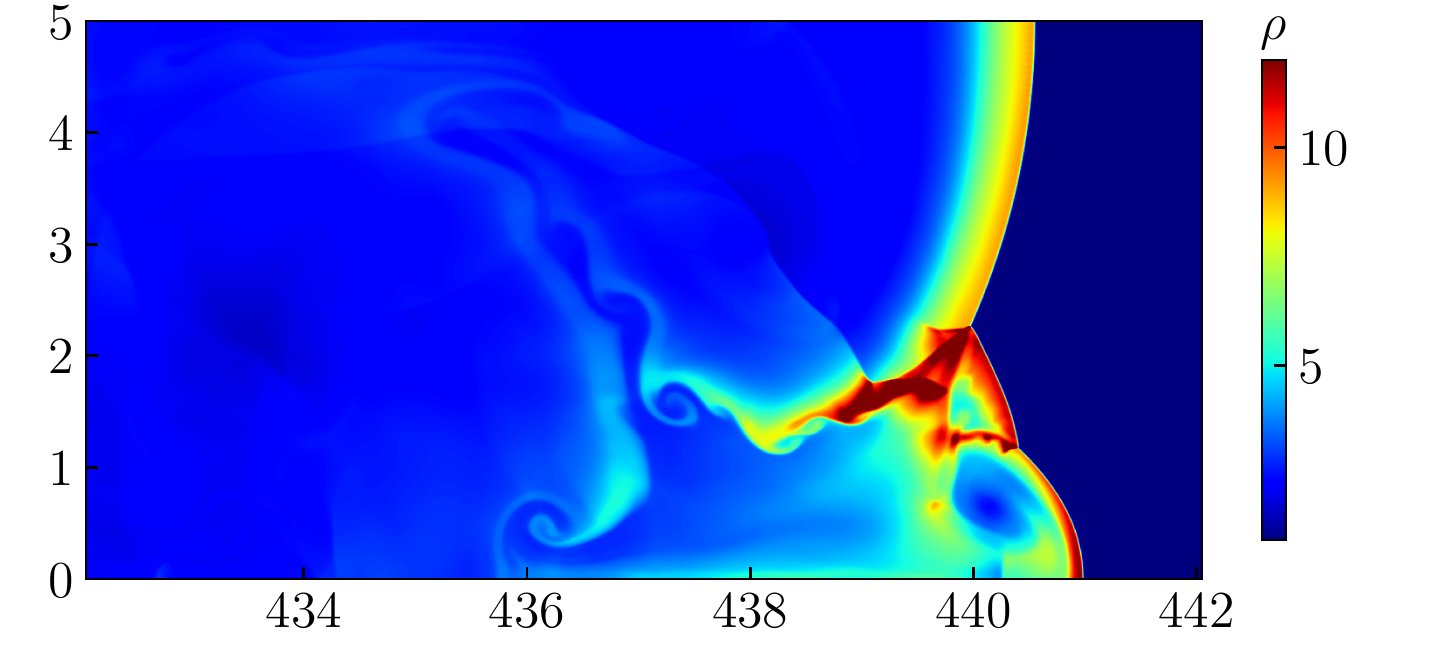} };
   \node at (30pt,90pt) {\textcolor{white}{d}};
        \node at (260pt,90pt) {\textcolor{white}{h}};
                  \node at (8pt,56pt) {\textcolor{black}{$y$}};
       \node at (110pt,2pt) {\textcolor{black}{$x$}};
                         \node at (238pt,56pt) {\textcolor{black}{$y$}};
       \node at (340pt,2pt) {\textcolor{black}{$x$}};
\end{tikzpicture}
\caption{Contour of density for: $\gamma=1.2$ a-d and  $\gamma=1.1$ e-h. Time sequences from a-d are:  $t=50.15$, $t=50.3$ , $t=50.6$, and $t=50.95$ and from e-h: $t=85.15$, $t=85.35$ , $t=86.1$, and $t=86.65$.   }
  \label{fig:q50-g-1.2-vs-1.1}
\end{figure*}

Further decreasing the polytropic index to $\gamma=1.2$ and  $\gamma=1.1$, the flow behind the leading front and the detonation structure become much more complex as depicted in Fig.\ \ref{fig:q50-g-1.2-vs-1.1}.  Like in Fig.\ \ref{fig:q50-g-1.66-vs-1.33}, the time sequences  starts when the main triple point, $tp$ is about to reflect from the bottom symmetrical plan\textcolor{black}{e}. The vortex structure clearly seen in Fig.\ \ref{fig:q50-g-1.2-vs-1.1}a is triggered from the previous collision of the main triple point at the upper wall creating a bifurcation point, k1, on the Mach stem. As time evolves,  the main triple point collides with the bottom wall and gets reflected.  Immediately after the collision,  forward and backward jets are generated, see Fig.\ \ref{fig:q50-g-1.2-vs-1.1}b. The forward jet interacts with the newly created Mach shock and generates a kink, k2, on it, see Fig.\ \ref{fig:q50-g-1.2-vs-1.1}c. The backward facing jet moves away from the leading shock. Also noticeable is the compression wave that propagates laterally and towards the main front. This compression wave  which was not visible in the previous cases with higher $\gamma$ can influence the main front. The trace of the bifurcation point on the Mach stem, k1, becomes less and less visible during the process. At the fourth time sequence shown in Fig.\ \ref{fig:q50-g-1.2-vs-1.1}d, the rolling vortex associated \textcolor{black}{with} k1 detaches completely from the shock. The bifurcation  disappears as the reflected triple point moves (rtp) towards the upper wall. Readers are invited to see the video in the supplementary material for more details. 

For the lowest polytropic index used, $\gamma=1.1$, the Kelvin-Helymothz instabilities appear along the slip lines, see Fig.\ \ref{fig:q50-g-1.2-vs-1.1}e-h. Besides, the reflected triple-shock merges with the bifurcation point k1 creating a new complex of triple-shock\textcolor{black}{s} that propagates toward the upper wall. This is in contrast with the case where $\gamma=1.2$ in which the bifurcation point k1 disappears prior to the arrival of the reflected triple-shock.   

\begin{figure*}[!h]
\centering
\begin{tikzpicture}
  \node[above right] (img) at (0,0) {\includegraphics[clip,  scale=0.4,  trim= 4.5cm 13.3cm  0cm  5.6cm]{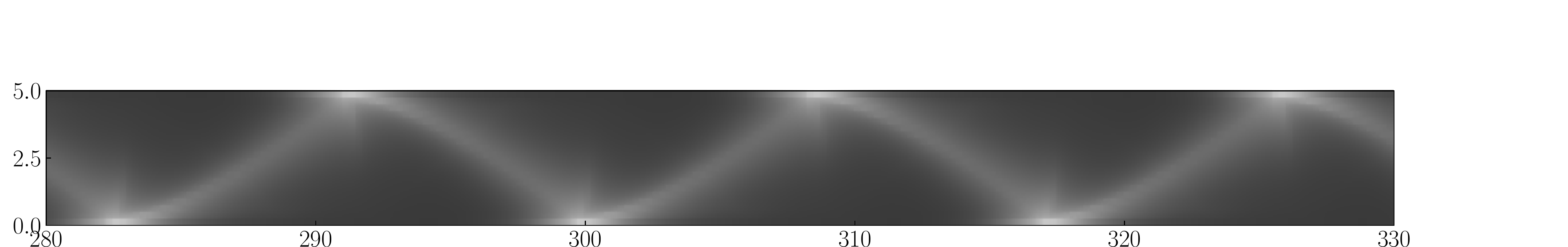}};
 \node at (35pt,40pt) {\textcolor{white}{a}};   
             \node at (0.5pt,32pt) {\textcolor{black}{$y$}};
       \node at (216pt,2pt) {\textcolor{black}{$x$}};   
 \end{tikzpicture}
 
  \begin{tikzpicture}
  \node[above right ] (img) at (0,0) {\includegraphics[clip,  scale=0.4,  trim= 4.5cm 10.4cm  0cm  10.6cm]{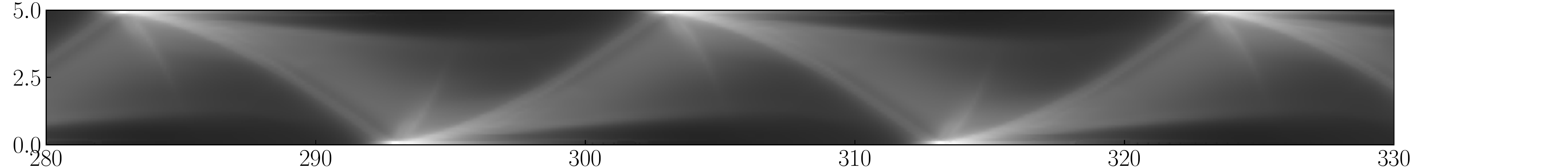}  };
  \node at (35pt,40pt) {\textcolor{white}{b}};
             \node at (0.5pt,32pt) {\textcolor{black}{$y$}};
       \node at (216pt,2pt) {\textcolor{black}{$x$}};
    \end{tikzpicture}
    
      \begin{tikzpicture}
  \node[above right ] (img) at (0,0) {\includegraphics[clip,  scale=0.4,  trim= 4.5cm 10.4cm  0cm  10.6cm]{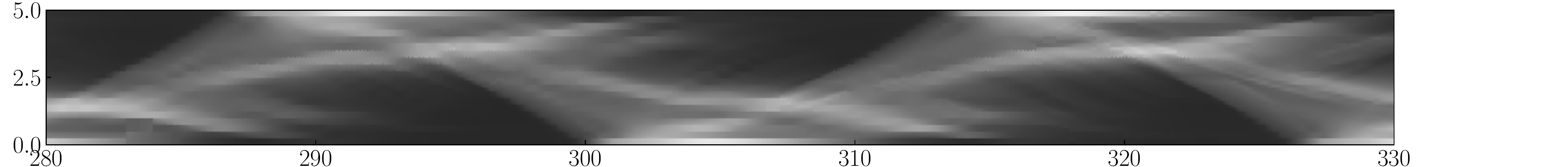}  };
  \node at (35pt,40pt) {\textcolor{white}{c}};
             \node at (0.5pt,32pt) {\textcolor{black}{$y$}};
       \node at (216pt,2pt) {\textcolor{black}{$x$}};

    \end{tikzpicture}
    
         \begin{tikzpicture}
  \node[above right ] (img) at (0,0) {\includegraphics[clip,  scale=0.4,  trim= 4.5cm 3cm  0cm  20.7cm]{Figs/shutter3.pdf}  };
               \node at (0.5pt,20pt) {\textcolor{white}{$y$}};

                       \node at (215pt,6pt) {\textcolor{black}{MERR}};

    \end{tikzpicture}
     
\caption{Numerical open shutter obtained by recording the location of the \textcolor{black}{maximum energy release rates (MERR)}  for : a) $\gamma=1.66$, b) $\gamma=1.33$,  and c) $\gamma=1.1$.} 
  \label{fig:q50-g-1.66-vs-1.1_openshutter}
\end{figure*}

The impact of the presence of the Mach stem bifurcation can be clearly seen in the numerical open shutter, obtained by recording the maximum energy release \textcolor{black}{rates} in each cell,  depicted in Fig.\ \ref{fig:q50-g-1.66-vs-1.1_openshutter}. For the high $\gamma$, the record is extremely clean with a single\textcolor{black}{-}headed configuration, see Fig.\ \ref{fig:q50-g-1.66-vs-1.1_openshutter}a. For the intermediate $\gamma$, shown in Fig.\ \ref{fig:q50-g-1.66-vs-1.1_openshutter}b, their \textcolor{black}{are} weak traces left by the weak bulge on the detonation front. And, for the lowest $\gamma$,  cell multiplication occurs, see  Fig.\ \ref{fig:q50-g-1.66-vs-1.1_openshutter}c.  
  
%
%
%
%
%


A close inspection of the reaction progress variable shows pockets of unburned gases that are created when the slip lines detach from the main front as in the experimental observations.  When the compressibility is enhanced, the forward jetting redistributes the progress  reaction variable behind the Mach stem thus modifying the reaction rate distribution near the Mach stem. The backward jets are also important as they are associated with the disruption of the pockets of unburned gases, see the accompanying video in supplementary materials.   
  \section{Convection mixing behind the main front}
\label{cnvectmix}

\begin{figure*}[!h]
\centering
\begin{tikzpicture}
  \node[above right] (img) at (0,0) {\includegraphics[clip,  scale=0.8,  trim= 0.cm 0cm  0.cm  0cm]{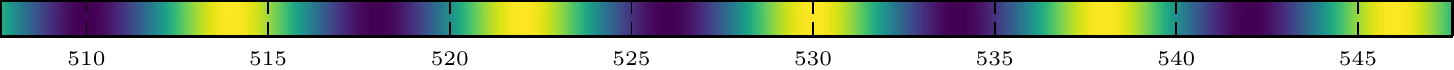}};
  \node at (30pt,15pt) {\textcolor{white}{a}};
 \end{tikzpicture}
 
 \begin{tikzpicture}
  \node[above right] (img) at (0,0) {\includegraphics[clip,  scale=0.9,  trim= 0.cm 0cm  0.cm  0cm]{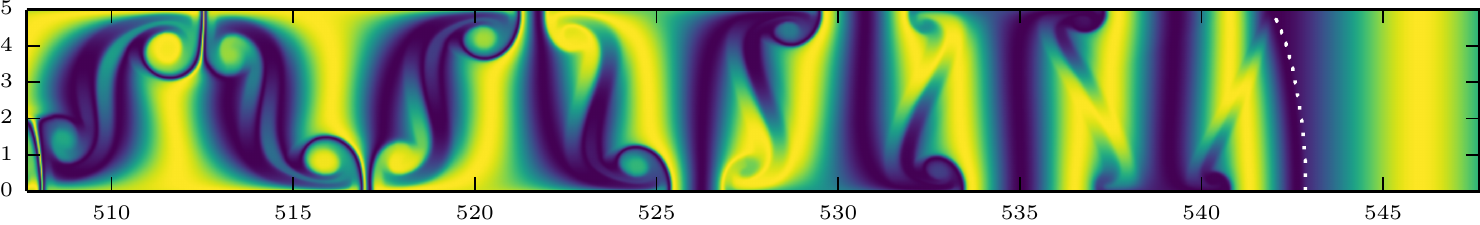}};
  \node at (20pt,40pt) {\textcolor{white}{b}};
            \node at (0.5pt,32pt) {\textcolor{black}{$y$}};
       \node at (202pt,2pt) {\textcolor{black}{$x$}};
 \end{tikzpicture}
 
  \begin{tikzpicture}
  \node[above right ] (img) at (0,0) {\includegraphics[clip,  scale=0.9,  trim= 0.cm 0cm  0cm  0cm]{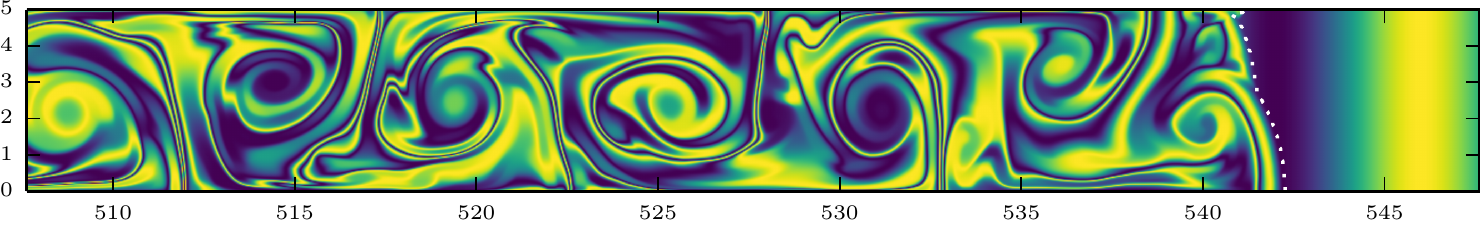}  };
  \node at (20pt,46pt) {\textcolor{white}{c}};
              \node at (0.5pt,32pt) {\textcolor{black}{$y$}};
       \node at (202pt,2pt) {\textcolor{black}{$x$}};

    \end{tikzpicture}
      
\caption{Plots of the Lagrangian flowfield coloured by $sin(\pi x_0/4)$ for:  a) the initial flowfield, b) $\gamma=1.66$ and c) $\gamma=1.1$ ; the white dotted-line is the leading shock front. }
  \label{fig:q50-g-1.66-vs-1.1_particles}
\end{figure*}

To further clarify the effects of the compressibility on the detonation structure, we investigate the convection mixing using a Lagrangian formalism. After a steady detonation structure is achieved in the channel, we introduced a passive scalar which stores the position of the gas ahead of the detonation front as follows:  
$$\displaystyle  \frac{D X(x,y)}{D t} = 0,$$ 
where $X(x,y)$ is the advected passive scalar storing the positions.  
The positions are coloured by taking $sin(\pi x_0/4)$. $x_0$ is the initial location of the advected position, see Fig.\ \ref{fig:q50-g-1.66-vs-1.1_particles}a.

Fig.\ \ref{fig:q50-g-1.66-vs-1.1_particles}b and c shows the advected passive scalar after the detonation front has passed.  While in the high polytropic index case the mixing is happening preferentially along the slip lines, in the lowest polytropic index case the mixing occurs almost everywhere in the channel as indicated by the rolling vortex motion, see Fig.\ \ref{fig:q50-g-1.66-vs-1.1_particles}b and c.        
%
%

\section{Statistical approach}
\label{staticalapproach}
In this section we adopt a global approach to quantify the effects of the compressibility on the reaction front. We extract the one-dimensional time dependent profiles of the progress variable at the bottom symmetrical plane. Then we set all the one-dimensional profiles in the shock-attached frame and applied a time averaging procedure, please refer to  \cite{sow_mean_2014}  for more details.

\begin{figure*}[!h]
\centering    
\begin{tikzpicture}
\node[above right] (img) at (0,0) {\includegraphics[clip, scale=0.4,  trim= 1.05cm 0.35cm  0cm 1.4cm] {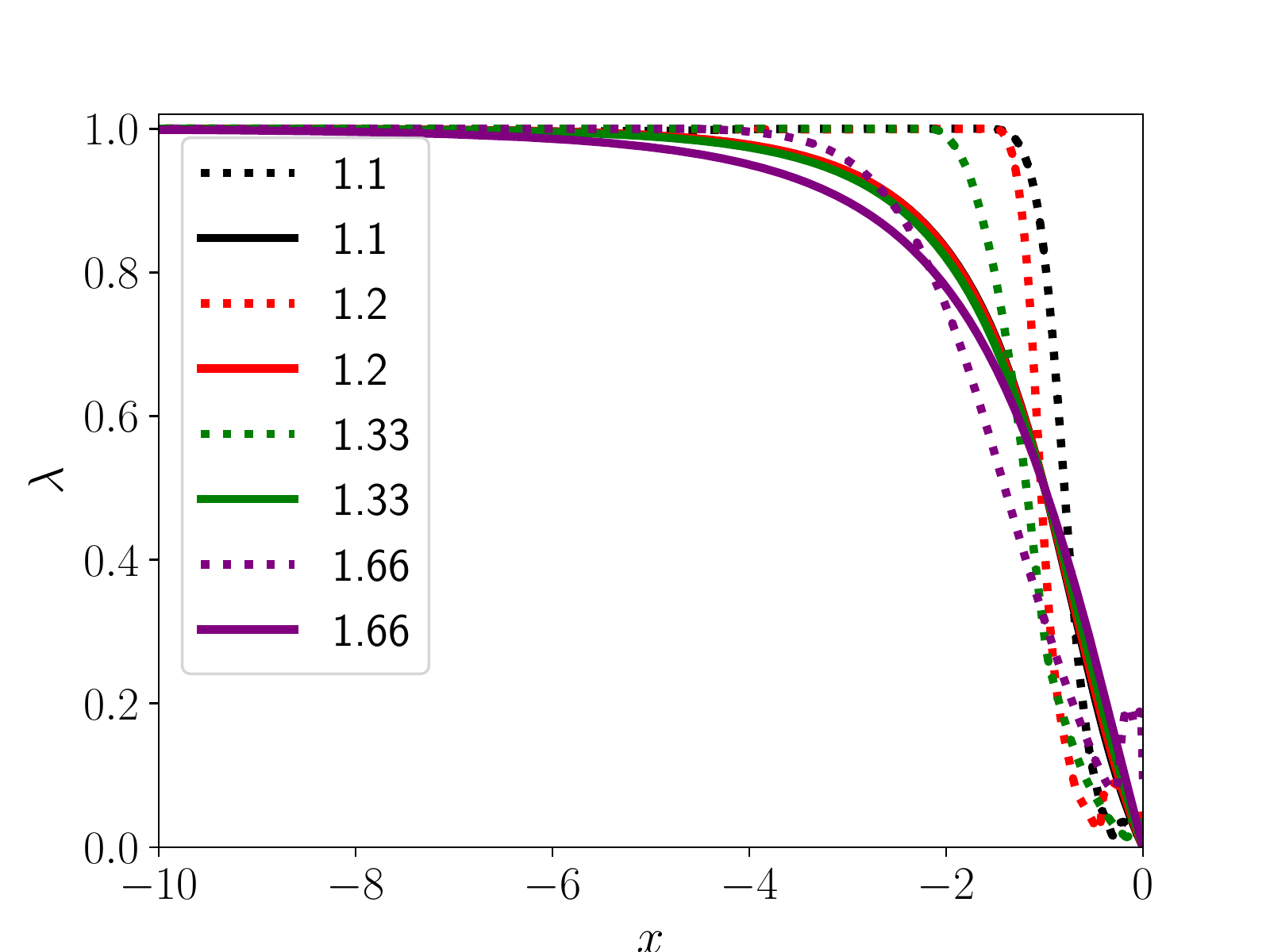} \includegraphics[clip, scale=0.36,  trim= 1.05cm 0.5cm  0cm 1.4cm] {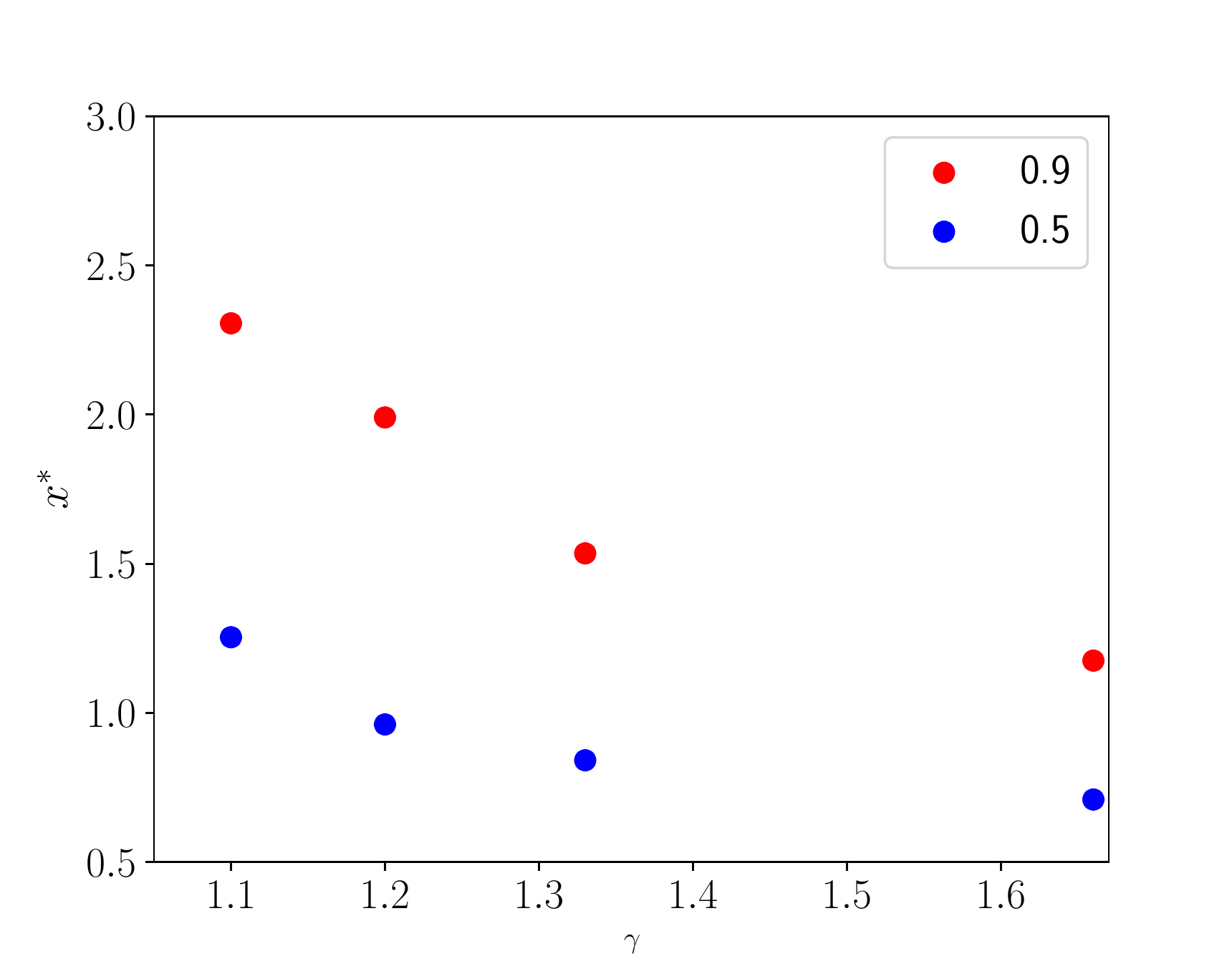}};
\node at (88pt,2pt) {\textcolor{black}{$x_{sh}$}};
 \node at (1pt,70pt) {\textcolor{black}{$\overline \lambda\;\;$}};
 \node at (260pt,2pt) {\textcolor{black}{$\gamma$}};
 \node at (170pt,70pt) {\textcolor{black}{$x^*$}};
 \node at (30pt,25pt) {\textcolor{black}{$a$}};
\node at (220pt,25pt) {\textcolor{black}{$b$}};

\end{tikzpicture}
\caption{Profiles of Reynolds averaged reaction progress variable in the shock attached frame (a). $x^* = \Delta_{\textrm{ZND}}/\Delta_{\textrm{cell}}$ as a function of $\gamma$ (b); $\Delta_{\textrm{ZND}}$ is the location at which $\lambda=\alpha$ in the ZND structure and  $\Delta_{\textrm{cell}}$ is the location at which $\overline \lambda=\alpha$ in the mean structure; $x_sh$ is the x coordinate in the shock-attached frame; $\alpha=0.5$ and $\alpha=0.9$; solid lines are for  \textcolor{black}{ZND} solutions and dotted lines are for averaged profiles.}
\label{fig:reactionzone}
\end{figure*}

Fig.\ \ref{fig:reactionzone} shows the obtained time averaged profiles. The steady ZND solutions are also plotted in Fig.\ \ref{fig:reactionzone} for comparison. The first thing to notice is that the steady ZND structure is weakly affected by $\gamma$.  This is due to the fact that we imposed the activation energy with respect to the post-shock state to be the same. The impact of the compressibility on the steady ZND reaction zone is negligible. For the mean reaction zone, however, the compressibility impacts significantly the reaction zone. As the compressibility is increased, most of the fresh matter is consumed closer to the leading shock, see Fig.\ \ref{fig:reactionzone}a. In the most compressible case, the mean reaction length is almost $2.4$ times smaller than the steady ZND reaction length. This is due to the fact that the increased compressibility tends to compress the fresh gases closer to the shock.          

\section{Mechanism of shock-bifurcation}
\label{comparisoninertshock}
  
The question we aim to address in this section is: {\it{ whether the \textcolor{black}{Mach stem  bifurcation due to jetting in detonations is a purely gasdynamics} driven problem?}}
A recent study \cite{lau-chapdelaine_viscous_2019} investigating the jetting and Mach stem bifurcations in inert shock reflections,  accurately defined  the bifurcation boundary between Mach stem bifurcations and no bifurcations. The governing parameters in \cite{lau-chapdelaine_viscous_2019} were the Mach number of the incident shock and the polytropic index. A fixed reflection angle of $30^{\circ}$, representative of shock reflection at triple point shock collisions in experiments, was used.  

\begin{figure}[!h]
\centering    
\begin{tikzpicture}
\node[above right] (img) at (0,0) {\includegraphics[clip, scale=0.6,  trim= 1.25cm 0.45cm  0cm 1.42cm] {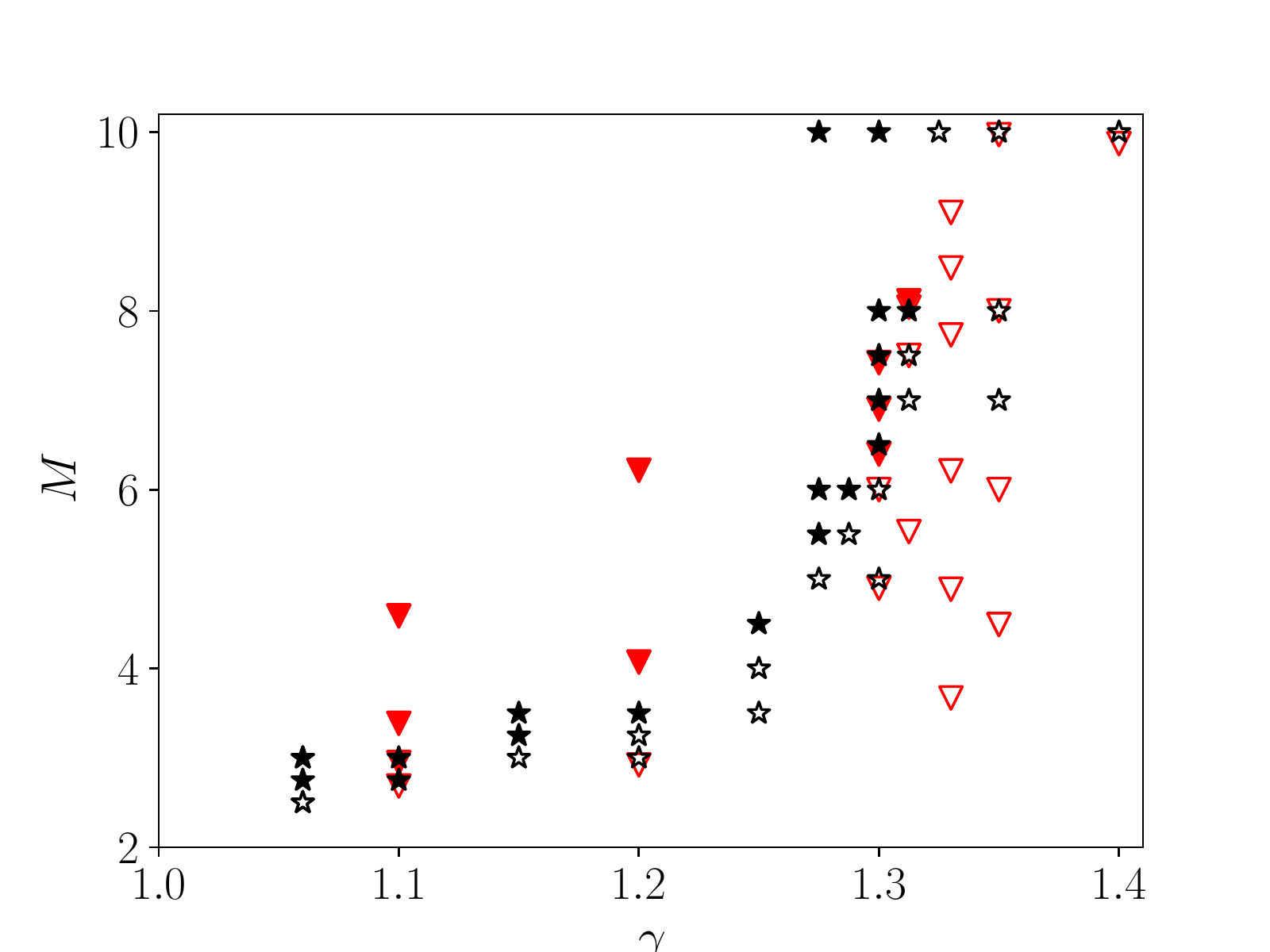}};
\node at (123pt,0pt) {\textcolor{black}{$\gamma$}};
 \node at (1pt,100pt) {\textcolor{black}{$M$}};
\end{tikzpicture}
\caption{Bifurcation domain: \textcolor{black}{full} symbols correspond to the bifurcating cases and \textcolor{black}{empty} symbols to the none bifurcating cases; \textcolor{black}{stars} are for inert shock simulations and triangles for detonation simulations; $M$ is the Mach number. \textcolor{black}{The Mach number used to construct the triangle symbols is the Chapman-Jouguet Mach number which is based on the initial  thermodynamic state and the heat release.}}
\label{fig:bifurcation}
\end{figure}

The findings are reported in Fig.\ \ref{fig:bifurcation}.  Filled and empty stars represent the simulations with bifurcation and without bifurcation, respectively.   The Mach number required for bifurcation increases when the isentropic index is increased up to $\gamma \approx 1.3$. Above this cut-off, no bifurcation was observed for Mach numbers lower than $7.8$.\\
We conducted detonation simulations using mixtures with same $E_a/RT_s$ to compare our results with the inert shock reflection predictions. The results are also reported  in \mbox{ Fig.\ \ref{fig:bifurcation}}.  The full and empty triangles represent the simulations with and without bifurcations, respectively.\\ 
Remarkably, the inert shock model predictions agreed very well with the detonation results. The cut-off polytropic index predicted by the inert shock reflection model is in good agreement with the detonation simulations. Thus, shock bifurcation in detonations \textcolor{black}{resulting from jetting appears to be} mainly a gasdynamic driven process.  
  
\textcolor{black}{Notice that today the  body of  work on experimental detonations in very low gamma mixtures is very limited. For instance, Imbert et al. \cite{imbert2005detonation} performed experimental detonations in stoichiometric n-heptane/oxygen and n-heptane/oxygen/argon mixtures at initial pressures between $2$ kPa to $7.5$ kPa. For n-heptane/oxygen mixture at $p_0=2$ kPa and $T_0=300$ K the post-shock frozen gamma is estimated to be $1.129$. At the same initial conditions with an argon molar fraction of $20\%$,  the post-shock frozen gamma increases to $1.153$.  A way to achieve gamma $1.1$ would be to use a rich mixture of n-heptane/oxygen. For example, in ‘C$_7$H$_{16}$:1.5 O$_2$:11’ at $p_0=2$ kPa and $T_0=300$ K, the post-shock gamma is $1.103$.}

Note that in the simulations conducted here,  constant specific heats have been employed. In real mixtures, however, heat capacities are temperature and composition dependent.  Future investigations will be dedicated to the effects of varying heat capacities. The role of the molecular diffusion phenomena on the mixing and the unburned pockets of gases need also further investigations.  \textcolor{black}{Furthermore,  the present study used a single couple of activation energy (with respect to the post shock state) and chemical heat release,  therefore wether the current findings can be generalized to all detonations needs additional investigation.}

\section{Conclusions}
\label{conclusion}
  The present study clarified the role of the compressibility in detonation structures. The study shows that the convection mixing is higher for low polytropic ind\textcolor{black}{icies} $\gamma$. Another important finding is that with decreasing $\gamma$ and higher compressibility, the fresh gases tend to be compressed closer to the leading shock causing the reaction zone to be shorter as compared to the steady ZND reaction zone. This study also shows conclusively that the Mach bifurcation \textcolor{black}{due to jetting} in detonations is mainly a gasdynamic driven phenomenon \textcolor{black}{for the thermochemical parameters considered.}

\section*{Acknowledgments}
\label{Acknowledgments}
The authors thanks NSERC for financial support through Discovery Grant to M.I.R and acknowledge the support of Compute Canada.  

\bibliography{referenceproci.bib} 
\bibliographystyle{elsarticlenumPROCI.bst}

\end{document}